\documentclass{aa}    
\usepackage{epsfig}
\def\h50{h$_{50}^{-1}${}}
\def\bj{b$_{\rm J}${}}
\def\kms{km~s$^{-1}$}
\def\ODM{`optical' dynamical mass{}}
\def\XDM{`X-ray' dynamical mass{}}
\def\LF{luminosity function{}}
\thesaurus{02(11.03.1, 11.03.4, 11.12.2) }

\def\R#1{{\mathrm{#1}}}		
\def\Eq#1{{Eq.~(\ref{e:#1})}}	
\def\Ep#1{{~(\ref{e:#1})}}	

\def\M#1{{\bf{#1}}}	
\def\T#1{{#1^{\bot}}}		
\def\d#1{{\R{d}{#1}}}		
\def\mdot{\!\cdot\!}		

\def\EQN#1{\label{e:#1}}        

\begin{document}
\title{The rich cluster of galaxies ABCG~85. IV. Emission line galaxies,
luminosity function and dynamical properties.
\thanks{Based on observations collected at the European Southern Observatory, 
La Silla, Chile}}

\author {
  F.~Durret \inst{1,2}
\and
  D.~Gerbal \inst{1,2}  
\and
  C.~Lobo \inst{3}
\and 
  C.~Pichon \inst{1,4,5}
}
\offprints{ F.~Durret, durret@iap.fr }
\institute{
	Institut d'Astrophysique de Paris, CNRS, 98bis Bd Arago, 
F-75014 Paris, France 
\and 
	DAEC, Observatoire de Paris, Universit\'e Paris VII, CNRS (UA 173), 
F-92195 Meudon Cedex, France 
\and
Osservatorio Astronomico di Brera, via Brera 28, I-20121 Milano, Italy
\and
Observatoire de Strasbourg, 11 rue de l'Observatoire, 67000 Strasbourg, France
\and
Astronomisches Institut, Universitaet Basel, Venusstrasse 7,
CH-4102 Binningen, Switzerland
}
\date{Received, 1998; accepted,}
\titlerunning{The rich cluster of galaxies ABCG~85. IV.}
\maketitle

\begin{abstract}

This paper is the fourth of a series dealing with the cluster of
galaxies ABCG 85. Using our two extensive photometric and
spectroscopic catalogues (with 4232 and 551 galaxies respectively), we
discuss here three topics derived from optical data. First, we present
the properties of emission line versus non-emission line galaxies,
showing that their spatial distributions somewhat differ; emission
line galaxies tend to be more concentrated in the south region where
groups appear to be falling onto the main cluster, in agreement with
the hypothesis (presented in our previous paper) that this infall may
create a shock which can heat the X-ray emitting gas and also enhance
star formation in galaxies.  Then, we analyze the luminosity function
in the R band, which shows the presence of a dip similar to that
observed in other clusters at comparable absolute magnitudes; this
result is interpreted as due to comparable distributions of spirals,
ellipticals and dwarfs in these various clusters. Finally, we present
the dynamical analysis of the cluster using parametric and
non-parametric methods and compare the dynamical mass profiles obtained
from the X-ray and optical data.


\end{abstract}

\section{Introduction}\label{intro}

As the  largest gravitationally bound systems in  the  Universe, clusters of
galaxies have attracted much interest since the  pioneering works of Zwicky,
who evidenced the  existence of dark matter in  these objects, and later  of
Abell (1958), who achieved the first large  catalogue of clusters.  Clusters
of  galaxies are  now   studied  through various   complementary approaches,
e.g. optical imaging and spectroscopy,  which allow in particular to  derive
the distribution and kinematical properties  of the cluster galaxies, and to
estimate  the luminosity function, and  X-ray  spectral imaging, which gives
informations  on the physical  properties of the  X-ray  gas embedded in the
cluster,  and with  some hypotheses can  lead  to estimate the total cluster
binding mass.

As a complementary approach to large cluster surveys at small redshifts such
as the ESO  Nearby Abell  Cluster Survey  (ENACS, Katgert et  al.  1996), we
have  chosen to  analyze  in detail  a few  low-z  clusters of  galaxies, by
combining   optical  data (imaging and spectroscopy    of a  large number of
galaxies)  and   X-ray data   from   the  ROSAT archive. We    present  here
complementary results on ABCG 85, which our group has already analyzed under
various aspects (see references below).

ABCG 85 has a redshift of z$\sim$0.0555, corresponding to a spatial scale of
97.0~kpc/arcmin  (for  H$_0=50$  \kms Mpc$^{-1}$, value  that   will be used
hereafter, together with q$_0$=0). Its  center  is defined hereafter as  the
center of   the  diffuse X-ray   component: $\alpha  _{\rm  J2000}=$~0$^{\rm
h}$41$^{\rm mn}$51.9$^{\rm s}$, $\delta _{\rm  J2000}=$~$-$9$^{\circ}$18'17"
(Pislar et al. 1997).  A wealth of data is now available for this cluster: a
photometric   catalogue of 4232 galaxies obtained    by scanning a \bj\ band
photographic plate in a square region $\pm 1^\circ$ (5.83 Mpc at the cluster
redshift) from the cluster center, calibrated with  V and R band CCD imaging
taken in the very center (Slezak et al.  1998) and a spectroscopic catalogue
of 551  galaxies  in a roughly  circular  region of 1$^\circ$ radius  in the
direction  of  ABCG 85,  among  which 305 belong to   the cluster (Durret et
al. 1998a). As  discussed  in  our previous papers    (Pislar et al.   1997,
Lima-Neto et al. 1997, Durret et al. 1998b),  there exists in fact a complex
of  clusters ABCG 85/87/89 in this   direction. In X-rays,  ABCG  85 shows a
homogeneous body, onto which are  superimposed various structures: an excess
towards  the north-west and south-west,  a south region  superimposed on it,
and several blobs  forming a   long filament  towards  the south-east;   the
velocity data confirm the    existence of groups and  clusters  superimposed
along the line of sight (see a complete  description in Durret et al. 1998b)
and show that this X-ray filament seems to be made of blobs falling onto the
main cluster.

We present here results on ABCG 85 which  may have cosmological implications
on  the  formation of galaxies,  clusters  and large   scale structures. The
properties of emission versus  non-emission line galaxies will  be discussed
in section \ref{emiss}  and compared   to  recent results on emission   line
galaxies in clusters by Mohr et al.  (1996) for ABCG 576 (redshift z=0.038),
Biviano  et   al. (1997)  for   the   large ENACS sample    at low  redshift
(0.035$<$z$<$0.121), and  Carlberg et  al.  (1996) for   the CNOC sample  at
higher redshift  (0.1709$<$z$<$0.5466).  The cluster luminosity  function in
the  R band  will  be derived in section~\ref{fdl}   and its  shape will  be
compared to     that   found in    other  clusters.    We   will  present in
section~\ref{dyn} the dynamical properties of the cluster, by estimating the
dynamical mass  from optical data with  various methods and  comparing these
results   with  the  dynamical    mass derived  from   X-ray  data. Finally,
conclusions will be drawn in section~\ref{discu}.

\section{Comparison of the properties of emission and non-emission
line galaxies}\label{emiss}

In a recent paper based on ENACS data, Biviano et al. (1997, the third paper
of the  series)  gave a good  review  of the  distribution and kinematics of
emission line (ELGs)   versus non-emission line  (NoELGs)  galaxies.   Their
approach is a statistical one, leading to  a general picture which fits well
in a current scenario of cluster formation.  On the  other hand, the present
study is devoted  to a specific cluster,   ABCG 85, with already  well known
general properties; the spirit  of this work  is therefore closer to that of
Mohr et al. (1996).

We  will  first compare  the  distributions of  ELGs versus  No\-  ELGs as a
function of various parameters.  As pointed out  by Mohr et al. (1996), this
roughly corresponds   to a separation into   gas-rich and gas-poor galaxies,
with  some contamination  of the gas-poor  sample expected.  It can  also be
considered as   a separation into  spiral  and non-spiral galaxies,  with an
underestimation of spirals, since not all spirals are ELGs.

The    numbers of   ELGs  and  NoELGs   in  our   sample   are 102  and  449
respectively. These numbers  are reduced to  33 and 272 respectively in  the
cluster velocity range, defined as the  13000-20000 \kms\ interval by Durret
et al.  (1998a).   This corresponds to   an  ELG fraction  of  0.11  in  the
cluster. Such a fraction is  in the range  derived by Biviano et al.  (1997)
for the  ENACS sample (0.08-0.12), but is  notably smaller than the value of
0.34 estimated by Mohr et al. (1996) for ABCG 576.

A  classification  of the ELGs  belonging  or not  to ABCG  85  based on the
equivalent  widths of   the main emission   lines will   be performed  in  a
forthcoming paper. The possible   existence and influence  of  environmental
effects on the presence and level of  activity in galaxies will be discussed
in that paper.

\subsection{Magnitude distribution}

\begin{figure}
\centerline{\psfig{figure=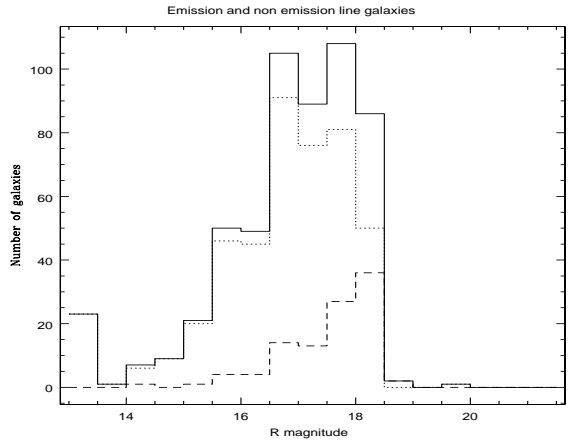,height=6cm,width=8cm}}
\caption[ ]{Histogram of the emission line (dashed line), non-emission
line (dotted line), and  total (full line)  numbers  of galaxies as  a
function of magnitude in the R band for all the spectroscopic sample.}
\protect\label{histoemnoem}
\end{figure}

\begin{figure}
\centerline{\psfig{figure=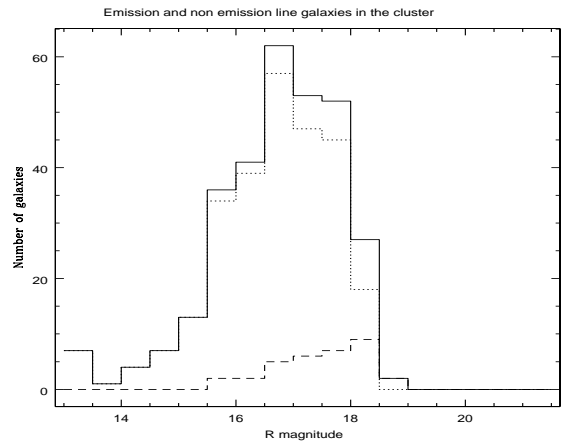,height=6cm,width=8cm}}
\caption[ ]{Histogram of the emission line (dashed line), non-emission
line (dotted line), and  total (full line)  numbers  of galaxies as  a
function of magnitude in  the R band  for galaxies in  the 13000-20000
\kms\ velocity range.}  \protect\label{histoemnoemin}
\end{figure}

\begin{figure}
\centerline{\psfig{figure=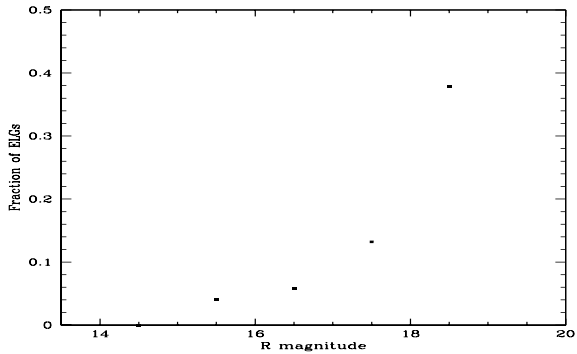,height=5cm,width=8cm}}
\caption[ ]{Fraction of emission line galaxies as a function of magnitude in
the R band for the galaxies with velocities in the 13000-20000 \kms\ velocity
range.}
\protect\label{fracemr}
\end{figure}

The  histograms of the distributions  of ELGs, NoELGs  and all galaxies as a
function of R magnitude are  displayed in Fig.~\ref{histoemnoem} for all the
galaxies in  our  redshift  catalogue.  A   similar  histogram is  drawn  in
Fig.~\ref{histoemnoemin}  for galaxies  belonging  to the  13000-20000 \kms\
velocity range. NoELGs are seen to be brighter than ELGs in both samples, as
in ABCG 576 (Mohr et al. 1996).

The fraction of emission line galaxies (defined in each magnitude bin
as the number of emission line galaxies divided by the total number of
galaxies) as a function of magnitude in the R band is displayed in
Fig.~\ref{fracemr} for cluster members. We note an increase of this
fraction with magnitude, as expected since for a given exposure time
redshifts are easier to obtain for emission line galaxies, which can
therefore be measured for fainter objects than for non-emission line
galaxies. Note however that this increase becomes steep for R$\geq$17,
i.e. even at magnitudes for which there is usually no problem to
measure absorption line redshifts, possibly suggesting that ELGs are
intrinsically fainter, though it is difficult to ascertain this result
since our redshift catalogue is by no means complete in the entire
region (its completeness is 85\% in a circular 2000 arcsec radius
region for R$\leq$18, then drops at larger radii, see Table 2 in
Durret et al. 1998a).  In order to quantify this effect, we calculated
the average luminosities for ELGs and NoELGs in the range
14.5$\leq$R$\leq$17.9. We find average luminosities corresponding to
magnitudes of 16.83 and 16.41 for ELGs and NoELGs respectively; the
standard errors on the luminosities give corresponding magnitude
ranges of [16.65, 17.04] and [16.35, 16.47], which do not overlap. The
Kolmogorov-Smirnov and Student tests indeed give probabilities that
these two samples originate from the same parent population of 0.0014
and 0.03 respectively, confirming that ELGs are indeed intrinsically
fainter that NoELGs.  This result is in agreement with that found on a
much larger sample by Zucca et al. (1997) of field galaxies.

\subsection{Spatial distribution}\label{espace}

\begin{figure}
\centerline{\psfig{figure=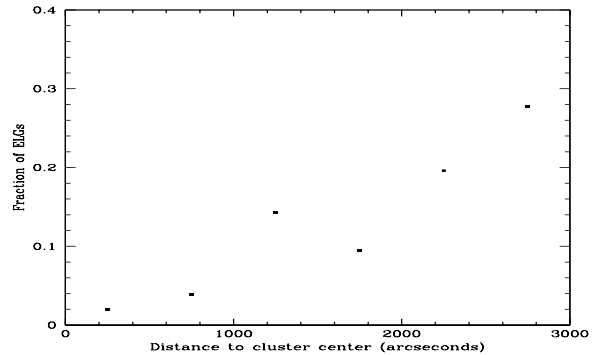,height=5cm,width=8cm}}
\caption[ ]{Fraction of emission line galaxies as a function of projected
distance to the cluster center for the galaxies in the 13000-20000 \kms\ 
velocity range.}
\protect\label{fracemd}
\end{figure}

The spatial distributions  of ELGs and NoELGs  are different for galaxies in
the velocity interval 13000-20000 \kms.   It is generally believed that ELGs
are more  frequent in the  outer regions  of the cluster   than in the inner
zones  (Biviano et al.  1997,  Fisher  et al.  1998). This  result is indeed
confirmed here  in   ABCG 85.   Fig.~\ref{fracemd}   shows  the fraction  of
emission line  galaxies (estimated as  the ratio  of the  number of emission
line galaxies to the total number of galaxies in concentric rings 500 arcsec
wide around the cluster center)  as a function of  projected distance to the
cluster center  for the galaxies   in the 13000-20000 \kms\  velocity range.
This fraction increases towards the outer regions of the cluster, implying a
difference in the spatial distributions of ELGs and NoELGs.

\begin{figure}
\centerline{\psfig{figure=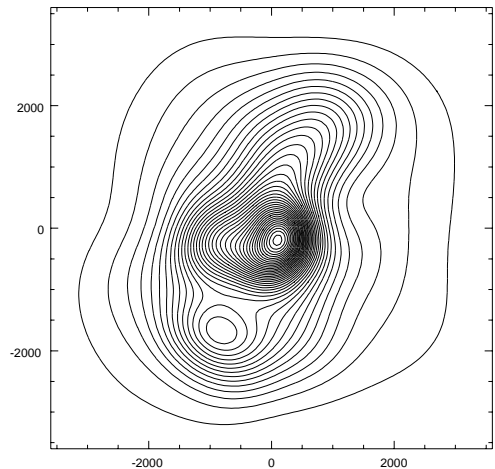,height=7cm}}
\caption[ ]{Adaptive kernel map of the spatial density distribution of
non-emission line galaxies. The axes are the positions relative to the
cluster  center (defined in the text)  in  arcseconds. North is to the
top and east to the left.}  \protect\label{kernoem}
\end{figure}

\begin{figure}
\centerline{\psfig{figure=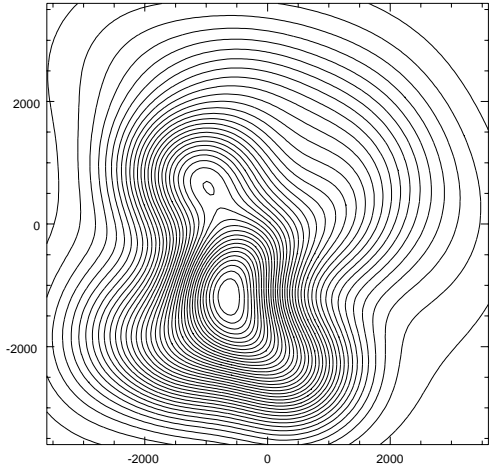,height=7cm}}
\caption[ ]{Adaptive kernel map of the spatial density distribution of
emission line galaxies. The axes are the positions relative to the cluster 
center (defined in the text) in arcseconds. North is to the top and east to 
the left.}
\protect\label{kerem}
\end{figure}

In  order to  visualize better the  spatial distributions  of both  kinds of
galaxies, we   have drawn adaptive  kernel maps  (e.g.  Pisani 1993)  of the
spatial density  distributions  of ELGs  and  NoELGs;  these are   shown  in
Figs.~\ref{kernoem} and  \ref{kerem} for all  galaxies  in the spectroscopic
sample,   disregarding  their  velocities.    The  distribution  of   NoELGs
(Fig.~\ref{kernoem})   is comparable to  that   derived from our much larger
photometric catalogue for all  galaxies, independently of spectral  features
and cluster  membership (see Fig.~1 in Slezak  et al. 1998): it is elongated
along PA$\sim 160^\circ$ and shows a  strong concentration around ABCG 85, a
secondary peak towards the   south east coinciding   with ABCG 87  (see  for
example Table 2 in Durret et  al.  1998b) and an  enhancement roughly at the
position  of  ABCG   89  to  the north   west  (as discussed  by   Durret et
al.  1998b). The  galaxy distribution  of  ELGs  (Fig.~\ref{kerem}) is quite
different: its  peak does not coincide  with ABCG  85, but  is close  to the
position   of   ABCG 87;  it   shows  a    weak secondary  maximum  in   the
north-northeast  direction  and  elongations  along  several  PAs, all quite
different from 160$^\circ$.

\subsection{Velocity distribution}\label{vitesse}

\begin{figure}
\centerline{\psfig{figure=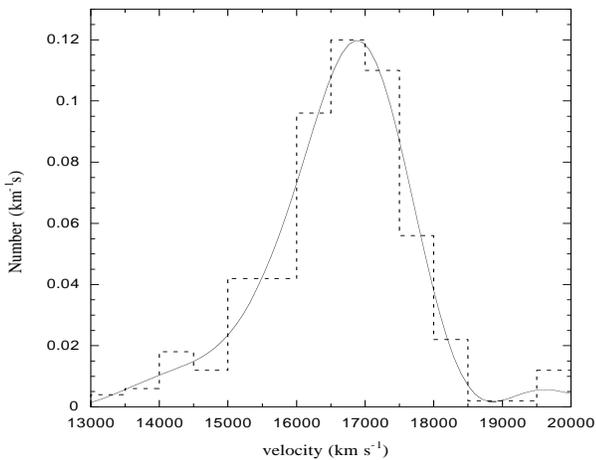,height=6.0cm,width=8cm}}
\caption[ ]{Histogram (dashed  line) and wavelet reconstructed density
distribution (full line) of the velocity of non-emission line galaxies
in the [13000-20000 km/s] interval.}  
\protect\label{wavevnoem}
\end{figure}

\begin{figure}
\centerline{\psfig{figure=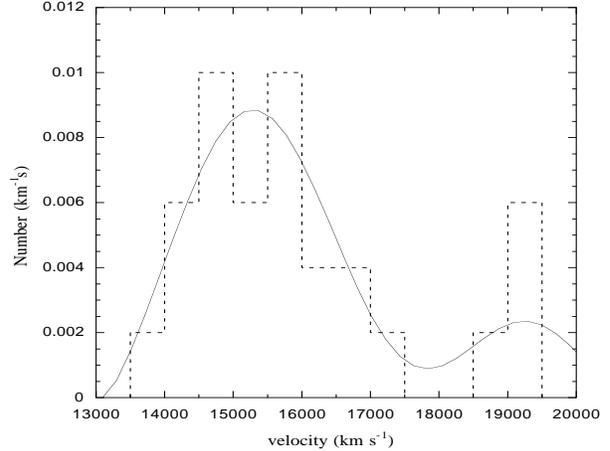,height=6.0cm,width=8cm}}
\caption[ ]{Histogram (dashed  line) and wavelet reconstructed density
distribution (full line) of the  velocity of emission line galaxies in
the [13000-20000 km/s] interval. }
\protect\label{wavevem}
\end{figure}

The mean and median velocities,   as well as the velocity  dispersions
are quite different for ELGs and NoELGs; the mean velocities are 15968
and 16627 \kms,  the median velocities are  15701 and 16732 \kms,  and
the velocity  dispersions are 1606 \kms\  and 1109 \kms\  for ELGs and
NoELGs respectively,  suggesting that the  morphology-density relation
is  coupled  with kinematic differences.  Biviano  et al. (1997) found
differences in the average  velocities of ELGs and  NoELGs at a  level
larger than 2$\sigma$ only for 12 clusters out  of their sample of 57;
their interpretation  was   that in   these 12  clusters  ELGs are   a
non-virialized  population falling onto the main  cluster.   In a much
smaller sample of 6  clusters, Zabludoff \&  Franx (1993) also found a
difference in mean velocity between spirals and early type galaxies in
3 clusters; on the other hand,  Mohr et al.  (1996) found that in ABCG
576  ELGs and  NoELGs had the  same  average  velocity, but with  ELGs
having a larger velocity dispersion, as in ABCG 85.

The velocity distributions displayed in Figs.~\ref{wavevnoem} and
\ref{wavevem} were obtained simultaneously using profile
reconstructions based on a wavelet technique and classical
histograms. We remind the reader that the features obtained with the
wavelet method are significant at a 3$\sigma$ level above the noise
(estimated at the smallest scale, see Fadda et al.  1998). While the
velocity distribution of NoELGs shows only one peak around 16800 \kms\
close to the mean or median previously given, that of ELGs shows a
peak at about 15300 \kms\ and a much smaller one around 19250 \kms.
Therefore, calculations of a mean value and of a velocity dispersion
for the total sample of ELGs in ABCG 85 do not characterize the ELG
velocity distribution properly. The mean (median) velocity for
galaxies belonging to the main ELG component (corresponding to
v$<$18000 km s$^{-1}$) is 15394 \kms\ (15411 \kms) and the velocity
dispersion is 900$\pm$190 \kms, much smaller than the value of 1606
\kms\ previously calculated. The second small peak includes five
galaxies, and has a mean (median) value of 19170 \kms (19320 \kms) and
a velocity dispersion of 320 \kms.  Three of these five galaxies are
very close both spatially and in velocity space, suggesting that they
are part of a physical group.  Notice that the velocity distribution
of NoELGs is not gaussian.  A more detailed velocity analysis of the
total sample of galaxies can be found in Durret et al. (1998b).

Furthermore, it is interesting to note that the \textsl{minimum} of
ELG velocity density is obtained for $\simeq 17800$ \kms\ close to the
\textsl{maximum} for NoELG velocity density ($\simeq 17000$ \kms); the
two samples are therefore different in velocity space, as confirmed by
statistical tests (see Table~1 and section 2.4).

The fraction of ELGs seems to increase with velocity. It is about 0.12
within  the cluster  velocity  range, and  increases  to 0.25-0.33 for
background  objects (those  with  velocities larger than 20000  \kms),
with  an extreme value  of 0.45  in the last  velocity bin (velocities
larger than 60000 \kms). Such variations are obviously at least partly
due to a selection effect (the  redshifts of faint background galaxies
are easier  to measure  for  emission  line  than for absorption  line
spectra), but are difficult to interpret because of the incompleteness
of our velocity catalogue at large radial distances.

\subsection{Are the ELG and NoELG properties significantly
different?}

As shown above, the spatial, magnitude and velocity distributions of
ELGs and NoELGs are different.  In order to quantify statistically
these results, we have tested the null hypothesis which would assume
both samples to be issued from the same parent population, both in
velocity space (v) and spatially (right ascension $\alpha$,
declination $\delta$ and projected distance to the cluster center D).

The statistical tests used are  the unpaired comparison t-test and the
Kolmogorov-Smirnov (K.-S.) test. The former is based on the comparison
of the means of  both samples  (ELGs and  NoELGs). Since we  have seen
that the velocity  distribution of  ELGs is  bimodal,  we have applied
this  test    in  two   velocity ranges:   (A)     [13000-20000 km/s],
corresponding   to  the   entire   cluster  velocity   range,  and (B)
[13000-18000 km/s],   where   the   second small  peak   in   velocity
distribution  of the ELGs  is eliminated, while the NoELG distribution
is not strongly affected.

\begin{table}[]
	\centering
	\caption{Statistical tests on the properties of ELGs and NoELGs}
	\begin{tabular}{l r r r r}
			\hline
			Probability of & v~~ & $\alpha$~ & $\delta$~~ & D~~ \\
			Null hypothesis &  &  &  &   \\
			\hline
			\hline
			Student (A) & 0.0074     &  &  &   \\
			Student (B) & $< 0.0001$ &  &  &   \\
			K.-S. (B)   & 0.0004     & 0.77 & 0.0490 & 0.0066  \\
			\hline
	\end{tabular}
	\label{statistique}
\end{table}

The results are shown in Table  \ref{statistique}, indicating that the
null hypothesis  is rejected  with  a high  degree of confidence. This
result   is  confirmed by the  non-parametric   K.-S. test,  which was
applied to the four characteristic properties of both samples.  Except
for  the   $\alpha$ variable,   all  the other  quantities   give weak
probabilities for the null hypothesis.

\subsection{Physical interpretation}

We have previously shown (Durret et al. 1998b) that ABCG 87 is made of
several small groups falling onto the main body of ABCG 85.  This
picture also allows us to give a general explanation of the ELG
properties described above.  The density-morphology relation (e.g.
Adami et al.  1998a and references therein) shows that the less dense
a region, the larger the rate of late type galaxies.  Spirals, and
consequently ELGs (which are mainly spirals) would therefore tend to
avoid the central region of ABCG 85.  Since groups of galaxies are
less dense, spirals tend to be more numerous in the ABCG 87 region
(both in space and in velocity space, see sections \ref{espace} and
\ref{vitesse}).  Moreover, the arrival of these groups onto ABCG 85
probably creates shocks, which leads the temperature of the X-ray
emitting gas to increase, as indeed observed in the ASCA temperature
map obtained by Markevitch et al. (1998). The shocks induced by the
merging of groups into the main cluster may well trigger star
formation in gas-rich spiral galaxies and account for the increase in
the number of ELGs in the ABCG 87 region (assuming that the general
identification of spirals with ELGs is valid).  Such a picture is
consistent with numerical simulations (e.g. Bekki 1998), in which
merging phenomena in clusters trigger star formation, and therefore
enhance the numbers of ELGs in merging regions.

The velocity dispersion in the main  ELG density peak is high (900$\pm
190$ \kms). In the general picture described above, the large velocity
dispersion found   for ELGs can  be explained  as   resulting from the
convolution of   the  velocity  dispersion in  each    blob (typically
$\simeq$300 \kms) with the velocity of each blob.  However, due to the
small number of ELGs,  it is difficult to  show this directly from the
ELG data.

In their interesting statistical analysis of the properties of ELGs in
nearby clusters based on the ENACS data, Biviano et al.  (1997)
emphasize some results, in particular the fact that ELGs appear to
avoid the central regions of clusters. They propose a schematic model
with two types of components, one with a velocity offset relative to
the average cluster velocity and a fairly small velocity dispersion,
and the other with no velocity offset and a large velocity dispersion.

These properties, combined with others, suggest that ELGs are falling
into the central region without having been previously in it. Such a
result is also found by Carlberg et al. (1996) for a sample of about
15 clusters with redshifts between 0.17 and 0.55.  Notice that the
larger velocity dispersion of ELGs compared to NoELGs can at least
partly be due to the difficulty of separating various velocity
subsamples; it is only in the case of large amounts of data and when
the distribution is clearly asymmetric that it is possible to improve
the analysis, as in our case. When a particular cluster is studied in
detail, one of the two types of components prevails. Infall is
generally not spherically symmetric, because it occurs preferably
along filaments (van Haarlem \& van de Weygaert 1993, West 1994); this
is the case in ABCG 85.

\section{Luminosity function in the R band}\label{fdl}

The study of luminosity functions  allows to give constraints not only
on the cluster galaxy content, such as  the relative abundances of the
various galaxy types,    but  also on  larger  scale   properties.  In
particular,   environmental effects have  recently  been  shown to  be
important  in several  clusters;    in Coma,  for   example,  Lobo  et
al. (1997) have shown that the faint end of the luminosity function is
steeper in  the cluster  than in  the   field, except in  the  regions
surrounding  the two large central galaxies.   This was interpreted as
due to the fact that each of these two galaxies is  at the center of a
group falling on  to the main cluster;  these  groups tend to  accrete
dwarf galaxies, and as a result  the luminosity function is flatter in
these regions.

We will discuss here the properties of the bright part of the luminosity
function of ABCG 85 in the R band. 

\subsection{Description of the available samples}

In order to  estimate the luminosity function,  we can use either  our
redshift catalogue or our CCD imaging catalogue.

The redshift catalogue covers a roughly circular region of 1$^\circ$
radius around the center of ABCG 85; it is the shallowest one: its
completeness is 82\% for R$\leq$18 in a circle of 2000 arcsec diameter 
around the cluster center (302 galaxies in this region). We will limit
our analysis to this region hereafter.

The CCD imaging catalogue in the  V and R  bands was obtained from 10 minute
exposures in each band, in a small region in  the center of the cluster (see
Fig.~4 in Durret et al. 1998a), covering an area of  246 arcmin$^2$; 381 and
805 galaxies were detected in the V and R bands respectively.

A photographic  plate catalogue  (4232 galaxies) has   also been obtained by
scanning a plate  in   the b$_{\rm J}$  band  (Slezak  et al.  1998).  It is
complete  down to  b$_{\rm  J}$=19.75  in a  $2^\circ\times  2^\circ$ square
region; however, since  it is shallower than the  CCD imaging catalogue,  we
will not use it here.

Background counts  were kindly made  available to us in  the R band from the
Las Campanas survey  (LCRS) by H.~Lin (see  Lin et al.   1996)  and from the
ESO-Sculptor    survey (ESS)  by  V.~de   Lapparent  and collaborators  (see
e.g. Arnouts et al. 1997). Note  that the LCRS  is made in  a wide angle and
therefore has small error bars  in each bin, but  is limited to R$\leq$17.8.
The  ESS is meant   to reach  very deep   magnitudes in  a  small  beam, and
therefore its number counts are  small at relatively bright magnitudes  (for
17$<$R$<$18).

\subsection{The R band luminosity function}

We have  chosen to draw  the luminosity function  in the R band, because our
CCD  imaging catalogue is  deeper  in  R  than in  V.   In the bright   part
(R$\leq$18),  we will  use  galaxies with redshifts   in the cluster  range,
therefore     avoiding the problem   of    background subtraction. For these
galaxies, the R magnitude was estimated from  the photographic plate b$_{\rm
J}$ magnitude, as explained in Slezak et al. (1998).

\begin{figure}
\centerline{\psfig{figure=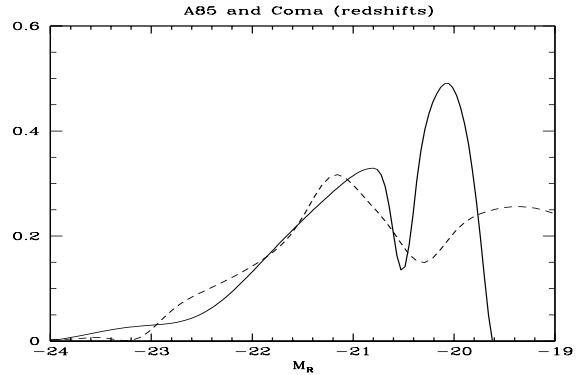,height=5cm,width=8cm}}
\caption[ ]{Wavelet reconstruction of the galaxy density as a function
of   absolute  R magnitude for  galaxies  belonging  to  ABCG  85 with
velocities  between 13000 and 20000  km/s (full line) and for galaxies
in Coma (dashed line). Arbitrary units are used to allow the direct 
comparison of both distributions.}  \protect\label{fdlz}
\end{figure}

Fig.~\ref{fdlz}  shows  a  wavelet   reconstruction of  the  distribution of
galaxies with velocities  in the   13000-20000  \kms\ velocity range  as   a
function of absolute R magnitude (with an adopted distance modulus of 37.6).
The wavelet reconstruction shows features significant at a level higher than
3$\sigma$.  In  the reconstruction  process,  we are   able to use  all  the
available scales    actually determined by  the  number  of galaxies  in the
sample.  However, we  are not interested by  phenomena at very small scales,
which anyway are rather noisy. We therefore excluded the two smallest scales
in our density profile reconstruction.   The resulting density profile shows
a dip for R$\simeq 17.1$, which corresponds to an absolute magnitude M$_{\rm
R}\simeq -20.5$.

We also derived  the luminosity function from the  R CCD imaging  catalogue,
which is complete to R$\sim$22, or M$_{\rm R}  \sim -15.6$, but in this case
it was necessary to  subtract a ``typical''  background contribution in this
band.  The ESS and the LCRS give consistent number  counts for the magnitude
bins  which they  have   in common (within  poissonian  error  bars and both
normalized to  the same area).    We  constructed a  background function  as
follows: for both surveys, we estimated the numbers of galaxies N$_{bg}$ per
square degree per magnitude bin; we merged  both surveys by considering that
the background  was represented by  the LCRS for R$\leq$17.8  and by the ESS
for R$>$17.8; we then fit several curves (power laws  or polynomials) to the
points  thus obtained. The  best  fit was reached for   a power law with the
following mathematical expression:

$ N_{bg} = 1.7544\ 10^{-23}\ {\rm R}^{19.758} $

\begin{figure}
\centerline{\psfig{figure=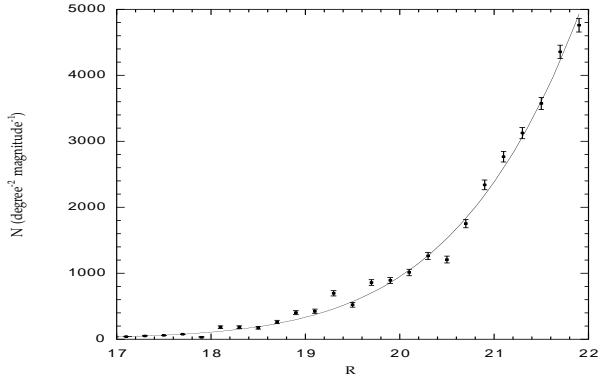,height=5cm,width=8cm}}
\caption[ ]{Best fit of the background contribution estimated from the
Las  Campanas redshift survey for  R$<$17.8 and  from the ESO-Sculptor
survey for R$>$17.8 (see text). } \protect\label{bg}
\end{figure}
\noindent
The background counts and fit are displayed in Fig.~\ref{bg}.

\begin{figure}
\centerline{\psfig{figure=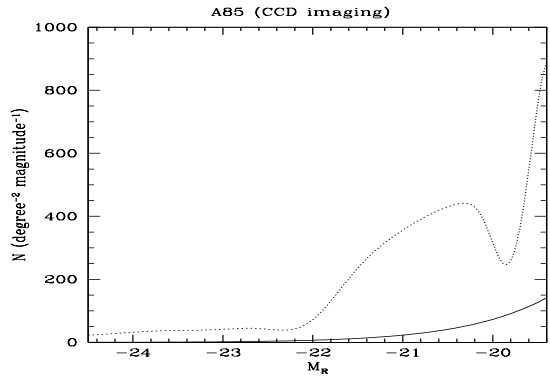,height=5cm,width=8cm}}
\caption[ ]{Wavelet  reconstruction  of  the  number of  galaxies  per
square  degree  and per  magnitude bin as    a function of  absolute R
magnitude  for   the  CCD imaging   sample  after  subtraction  of the
background  contribution (see  text). The   bottom line indicates  the
background contribution.}  \protect\label{fdlccd}
\end{figure}

Fig.~\ref{fdlccd} shows a wavelet reconstruction of the distribution
of galaxies derived from our CCD imaging catalogue after subtraction
of the background contribution as explained above.  The curve has
roughly the same shape as that displayed in Fig.~\ref{fdlz} for
galaxies with redshifts, but it is shifted by $\sim$0.6 magnitude,
with a dip now at M$_{\rm R}\sim -19.9$. As discussed in section 3.4,
it was not possible to draw the luminosity function for fainter
magnitudes because of the background subtraction problem.

\subsection{A dip in the luminosity function?}

\subsubsection{How real is the dip in the luminosity function?}

\begin{figure}
\centerline{\psfig{figure=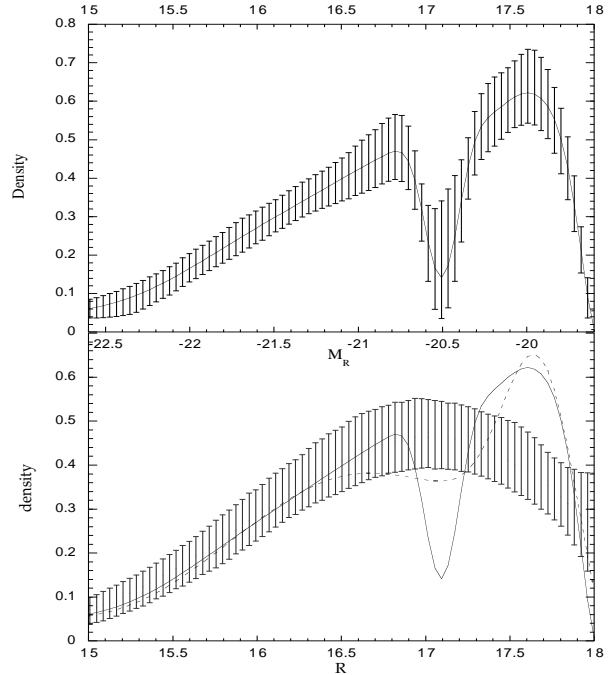,height=9cm,width=8cm}}
\caption[ ]{Top panel: Luminosity  function as in Fig. \ref{fdlz} with
error bars obtained by a bootstrap technique (see text). Bottom panel:
wavelet reconstructions of the luminosity functions at three different
scales;  error bars  are  obtained  by  a Monte-Carlo  technique  (see
text). The dashed line in the bottom panel shows the luminosity function
at a scale twice that of the full line.} \protect\label{fadda}
\end{figure}

As seen above, the wavelet reconstruction of the galaxy distribution
shows a dip. In order to illustrate the robustness of this result, we
have done two calculations.  First, we consider that our data are the
only available realization of a parent sample.  Therefore, the
bootstrap technique proposed by Efron (1979, 1982) seems well adapted
to estimate error bars.  We perform 1000 Monte-Carlo draws and do a
wavelet analysis on each of the 1000 draws.  We choose as limits to
the error bars the 10 and 90 percentiles of the distributions thus
obtained. These are shown on the top panel of Fig. \ref{fadda}. The
dip therefore appears to be statistically significant.  However, this
bootstrap technique gives too large a weight to the observed
realization; in particular, if a gap is present in the data, no draw
will be able to fill it.

We have  therefore applied a  second method. As a  first step, we have
wavelet reconstructed  the luminosity function  eliminating the  three
smallest scales.  The distribution obtained in this  way does not show
any dip. We have then performed 1000  Monte-Carlo draws following this
profile, and  again  have done a   wavelet analysis  on each  of these
draws.

The result of this method is shown in Fig.  \ref{fadda} (bottom
panel). The dip clearly appears outside the error bar region, implying
that the probability to obtain such a feature from such a parent
sample (devoid of dips) is smaller than 0.001; even the luminosity
function drawn at a larger scale (dashed line in Fig. \ref{fadda})
shows a shallower but still significant dip.

\subsubsection{Physical interpretation of the dip}

A comparable dip was found in the luminosity function of several
clusters.  We give in Table \ref{dip} the positions of the dips for R
band absolute magnitudes recalculated when necessary for a Hubble
constant of 50 km s$^{-1}$ Mpc$^{-1}$; luminosity functions drawn in
the B band have been shifted to the R band assuming B$-$R=1.7 for all
clusters except Virgo, a typical value for ellipticals, taken as the
dominant cluster population. For Virgo, where spirals are dominant, we
took B$-$R=1.4. No K-correction or Galactic absorption correction were
included, since this is only a rough comparison.

It is interesting to note that the dips in the luminosity functions
are found at comparable absolute magnitudes in all these clusters
within a range of only one magnitude. The only cluster that we found
in the literature having a dip at a significantly different absolute
magnitude is ABCG 496.

As  mentioned  above, the   dip  position derived from   the
redshift    catalogue differs  from that   derived   from the CCD  imaging
catalogue  in  ABCG  85. This  apparent discrepancy    is most likely
accounted for  by   the fact that  the latter   corresponds  to a much
smaller central region, and suggests that environmental effects 
modify the luminosity function in this cluster (see below).

These dips do not all seem to have the same width: the dip found in
the luminosity function of Shapley 8 is notably broader, while
shallower dips (or at least flattenings) are found in the luminosity
functions of Virgo and ABCG 963. However the methods used by these
various authors are quite different from ours; we have redone the
analysis described by Biviano et al. (1995) in the Coma cluster using
the wavelet reconstruction technique; the corresponding luminosity
function is displayed in Fig.~\ref{fdlz} and the dip has a shape
notably broader than that of ABCG 85.

\begin{table}[tbp]
\caption{Dip positions in clusters. ABCG 85(z) and (CCD): luminosity functions
derived from the reshift and CCD imaging catalogues respectively.}
\centering
\begin{tabular}{llcl}
\hline
Cluster name & Redshift & Dip position & References  \\
\hline
ABCG 85(z) & 0.0555 & -20.5 & This paper  \\
ABCG 85(CCD) & 0.0555 & -19.9 & This paper  \\
ABCG 496   & 0.0328 & -18.0 & Molinari et al. (1998) \\
ABCG 576   & 0.038  & -19.5 & Mohr et al. (1996) \\
ABCG 963   & 0.206  & -19.8 & Driver et al. (1994) \\
Coma       & 0.0232 & -20.5 & Biviano et al. (1995) \\
Shapley 8  & 0.0482 & -19.6 & Metcalfe et al. (1994) \\
Virgo      & 0.0040 & -19.8 & Binggeli et al. (1988) \\
\hline
\end{tabular}
\label{dip}
\end{table}

The above facts suggest that the bright  galaxy distributions in these
clusters have roughly comparable properties, but also that they differ
from  a cluster to another,  and even from one  region to another in a
given cluster. This is also  the case for  the relative abundances  of
galaxy types, which depend  on the local  density and/or on the global
properties of each cluster.

\begin{figure}
\centerline{\psfig{figure=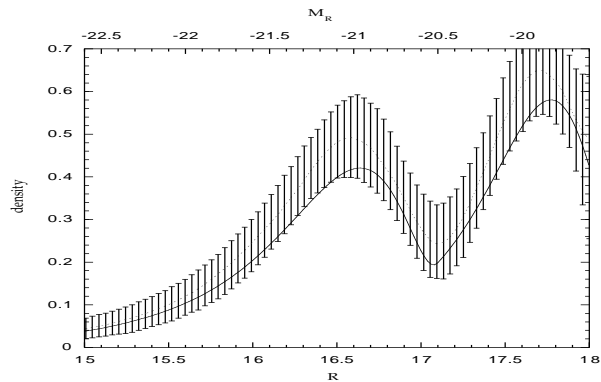,height=5cm,width=8cm}}
\caption[ ]{Simulated luminosity function. The  bold curve is obtained
by   the analytical method    described in the   text. The dashed curve 
corresponds to the median of Monte-Carlo draws. Error bars are
obtained  by   a Monte-Carlo    technique  as described   previously.}
\protect\label{toy2}
\end{figure}

In fact, a simple model based on the shapes of the luminosity
functions of the various galaxy types and on their relative
proportions (e.g. B\"ohm \& Schmidt 1995, Jerjen \& Tammann 1997, and
references therein) can roughly account for the dip in the luminosity
function.  

By using only three kinds of luminosity functions, and adjusting the
relative proportions of these three types of galaxies, it is easy to
recover a luminosity function with a similar shape to that observed.
As an example, such a toy-luminosity function is given in
Fig.\ref{toy2}; in this case, we have used:

- a Gaussian with $\mu_{R}=$15.8 
and $\sigma=$1.1, for the spiral luminosity function;

- a Gamma density (also called Erlang density): 
$$G(R) = \frac{\lambda^{\mu}}{\Gamma(\mu)} (R-R_{0})^{\nu -1} \exp
-\lambda (R-R_{0})$$ for ellipticals. This density function is
asymmetric and has been used by Biviano et al (1995) to describe the
luminous part of the Coma \LF; $R_{0}$ is a cut-off value (in the
example we have chosen $R_{0}=$17.2) the maximum is given by
$R_{max}=R_{0} - \frac{\mu -1}{\lambda}$;

- a power law to represent faint or dwarf galaxies, filtered by an
apodisation function to account for the incompleteness for high values
of R.

We can see from Fig.\ref{toy2} that although located at the good
position, the dip is broader than the ABCG 85 dip and not as deep, and
that the luminous part of the \LF\ is convex instead of concave.  In
fact, because the game is played with at least three functions, each
of them driven by 3 or even 4 parameters, we have too many degrees of
freedom and are able to modify these features in various ways.
However, the location of ellipticals relative to dwarfs is well
determined: the dip corresponds to the transition zone between
ellipticals and dwarfs. However, it has not been possible to play the
same game with two functions only.

In this hypothesis, the fact that the dip falls roughly at the same
absolute magnitude for at least seven clusters suggests that in the
dip region these clusters have comparable galaxy populations; however,
the fact that the various dips are not located exactly at the same
absolute magnitudes and have different widths raises the question of
the relative positions and densities of these various populations.
Lobo et al.  (1997) have shown that the slope of the faint luminosity
function varies with the local environment.  The less rich a cluster,
the more numerous are spiral galaxies; an increase in the number of
spirals will modify the luminosity function only around M$_{\rm R}\sim
-23$. The combination of these well established results leads to
confirm, as suggested by previous authors, that in general luminosity
functions depend simultaneously on type and on local density and/or
global properties (such as cluster richness; see e.g. Phillipps et
al. 1998).  However, our data is not complete enough to allow a
further analysis.

\subsection{A word about the faint end of the luminosity function}

We initially intended to analyze the faint end of the luminosity
function derived from our CCD imaging catalogue after subtracting the
background as described above. However, for M$_{\rm R}>-19.4$ the
luminosity function is found to decrease dramatically, rapidly
reaching negative values, although the data sample appears to be
complete up to R$\sim 22$. This implies that the background
contribution has been overestimated, i.e. the ``Universal'' background
counts as obtained in previous section (Fig. \ref{bg}) are not
representative of the background of ABCG 85.

Marginal evidence  for  the  existence of  a background  larger   than
expected from  statistical arguments has also been  found in Coma. Out
of 51 redshifts obtained for faint galaxies  (R$\leq 21.5$) in a small
region near the Coma cluster center,  at most five galaxies were found
to be cluster members (Adami et  al. 1998b), while the expected number
was $16\pm  11$. Due to  the small number  of redshifts involved, this
result is of course still preliminary; however, it raises the question
of  statistical background    subtraction, which   should  not be   as
universally accepted as it is now.

\section{Dynamical analysis}\label{dyn}

With the assumption that  the X-ray emitting gas  is isothermal and in
hydrostatic   equilibrium  with dark matter,  the   dynamical mass was
estimated as a function of distance to the cluster center by Pislar et
al. (1997) from X-ray (ROSAT PSPC) data (see  their Figure 13). Within
the X-ray image, which is limited to a distance of  1.3 Mpc around the
cluster center,  the  total  mass  estimated  is about  $3\   10^{14}$
M$_\odot$.

The basic physical picture behind this determination is that if the
cluster ABCG~85 is in equilibrium, then the velocities of any
subpopulation should reflect the mass distribution.  We have
calculated the ellipticity of the galaxy distribution using the
momentum method.  This provided us with a direction of the major axis
and the ratio $\epsilon$ of the small to large axes. We have then
changed the coordinates of the position of each galaxy by an
anamorphosis along the major axis: $X = \epsilon X$.  The galaxy
distribution then appears spherical. This defines a new radius $R$
which will be used in the following equations. This transformation
assumes that the cluster major axis is parallel to the plane of the
sky, in the prolate as well as the oblate cases.  We may then infer
the enclosed mass from the measured velocities of a tracer population.
Another possible method which is commonly used is to perform counts in
concentric circular rings. Our method is more accurate since the
galaxy count estimate is adapted to the geometry, but equation (2)
assumes spherical symmetry and is not fully correct. The other method
cumulates both defects.

Let us investigate the properties of the cumulative mass profile
derived from the different tracers when isotropy is assumed. This is
achieved via parametric and non parametric modeling.  The Appendix
gives a more detailed description of the non-parametric methods
involved.

\subsection{Data structure}
\label{s:data}

\subsubsection{Binning procedure}

Optimal inversion techniques should avoid binning while relying on
techniques such as Kernel interpolation.  We found here that for such
a sample and when assessing quantities which are two derivatives away
from the data, the Kernel introduces spurious high frequency features
in the recovered mass profile.  Binning the projected quantities on
the other hand allows us to control visually the quality of the fit.
We use floating binning which is defined as follows: for each galaxy
we find its p-nearest neighbors, and define a ring which encompasses
them exactly; the estimator for the density, $\Sigma_{\R{Ring}}$,
would be defined as p divided by the area of that ring.  For the
projected velocity dispersion squared, $\sigma _p^2$, we could sum over
the velocity squared (measured with respect to the mean velocity of
the cluster) of the p neighbors and divide by p; in practice a better
estimator, $\sigma _p^{2\star}$, accounting for velocity measurement
errors is given by
\begin{equation}
\sigma _p^{2\star}  = \left(
 \sum_{i \,   \in \, \R{Ring}} \frac{  v_i^2}{
\sigma _i^2} \right)\left/\right.  \left(
 \sum_{i \,   \in \, \R{Ring}} \frac{ 1}{
\sigma _i^2} \right)   \,, \EQN{ disp}
\end{equation}
\noindent
where $\sigma _i^2$ is the sum  of the error on the measured variance, 
$ \sigma _p^2$, and of the square of the measured error on the velocity $v_i$.
Bootstrapping is applied to estimate $\sigma _p^2$ while first
neglecting these measurement errors.  An estimate of the projected
energy density is given by $\sigma _p^{2\star}\ \times \Sigma
_{\R{Ring}}$.  In practice, binning over $10$ to $15$ neighboring
galaxies is applied, yielding estimates of the Poisson noise induced
by sampling.

\subsubsection{Bias and incompleteness}
\label{bias}

The sample is truncated in projected radius $R$.  Since generically,
truncation and deprojection will not commute, the estimation of the
cumulative mass profile arising from a truncated sample in projected
radius will be biased.  In physical terms, this follows because we
cannot distinguish between projected galaxies which are truly within a
sphere of radius $R$, and those which are beyond but happen to fall
along the line of sight.  Considerations about the physical properties
of the tracer may help reduce the confusion, but a bias remains in the
estimated mass when the sample is truncated in projected radii.
Extrapolation provides some means of correction.  Note that
extrapolation has a different meaning depending on what the true
profile is.  Specifically the boundary conditions (exponential
splines, edge spline, truncation at two or five times the last
measured radius) will make a difference in the recovered profile.
Since the completeness of our redshift catalogue decreases with
increasing radius, we restrict our analysis to the inner region of the
cluster within 1000 arcsec (1.62 Mpc at the cluster redshift), where
this catalogue is fairly complete (92\% for R$\leq 18$). In practice
we check that all mass estimates converge to the same total mass
within the error bars.

\subsection{Method }
\subsubsection{Jeans equation}

The equilibrium of  an isotropic stationary spherical galactic  cluster
obeys Jeans' equation:
\begin{equation}
M_{dyn}(<r)  =  r^{2} \frac{\d \psi}{\d r   } = -\frac{r^{2}}{\rho} \frac{\d
{(\rho \sigma_r^2) }}{\d r } \,, \EQN{mass0}
\end{equation}
where $\psi(r)$ is the gravitational potential generated by all the 
types of matter, i.e. stellar matter, X-ray emitting plasma and 
unseen-matter, $\rho(r)$ the density
 of  galaxies in   the  cluster and   $\sigma_r(r)$  the  radial velocity
 dispersion.   \Eq{mass0} can   be  applied   locally to  assess   the
 cumulative dynamical mass profile.

  The surface density  of galaxies  is  related to the density  via an
  Abel transform:
\begin{equation}
\Sigma(R) = \int\limits_{-\infty}^{\infty} 
 \rho(r) \d z   = 2 \int_{R}^\infty \rho(r)
\frac{r \d r }{\sqrt{r^2-R^{2}}} \equiv {\cal A}_{R}(\rho)\, , \EQN{Sigma}
\end{equation}
where   $\Sigma(R)$ is the projected   galaxy  density and $R$ the projected
radius as measured  on the sky. Similarly  the line of sight  velocity
dispersion  $\sigma_{p}^{2}$ is related   to  the intrinsic radial velocity
dispersion, $\sigma_r^{2}(r)$, 
via the \textsl{same} Abel transform (or projection)
\begin{equation}
\Sigma(R) \sigma_{p}^{2}(R)=2 \int_{R}^\infty \rho(r) \sigma_r^{2}(r)
\frac{r \d r }{\sqrt{r^2-R^{2}}} \equiv {\cal A}_{R}(\rho  \sigma_r^{2}) \, . 
\EQN{Sigmau2}
\EQN{epsilon}
\end{equation}
Note that $\Sigma(R) \sigma_{p}^{2} $  is the projected kinetic energy
density divided by three (corresponding  to one degree of freedom) and
$\rho(r) \sigma_r^{2} $  the kinetic  energy density  divided by three. 
Inverting Eqs. (3)-(4) 
into \Eq{mass0}  yields:
\begin{equation}
M(<r)    = -\frac{r^{2}}{{\cal  A}^{-1}_{r}(\Sigma)} \frac{\d
{{\cal A}^{-1}_{r}(\Sigma \sigma_{p}^{2})}}{\d r } \, . \EQN{mass}
\end{equation}  Therefore, assuming we have 
access to estimators for $\Sigma$ and $\Sigma \sigma_{p}^{2}$, the 
cumulative mass distribution follows.

\subsubsection{Algebraic Dynamical Mass  estimators}

The  Bahcall \& Tremaine (1981) mass estimator  for  test particles around a point
mass  assumes completeness and isotropy and is given by:
\begin{equation}
M( <  R_0)
\approx
 \frac{16}{G \pi}  \sum_{i  | R_i< R_0}   R_i  {v_i^2} \, .
 \EQN{BT}
\end{equation} 

\subsection{Parametric modelling}

In a nutshell, given that formally the inverse of 
 \Eq{Sigmau2} is 
  \begin{equation}
\rho \sigma_r^2 = {\cal A}^{-1}_{r}(\Sigma  \sigma_{p}^{2}) = - \frac{1}{\pi} 
\int_{r}^{\infty} 
\frac{\d{} \Sigma \sigma_p^2 }{d R}  \frac{\d R}{\sqrt{R^{2}-r^{2}}}
\,  \d R \, , \EQN{invAbel2}
\end{equation}
it is straightforward to construct parametric pairs of 
projected and deprojected fields. 

In order to describe the density profile of galaxies, and/or the
energy density, we have used various kinds of parametric forms:~

\begin{enumerate}
	\item A $\beta$-model: 

\begin{equation}
\rho(r) =  \rho_{0} \left(1+ \frac{R^{2}}{b^{2}}\right)^{-3 \beta/2} 
  \, , \EQN{beta}
\end{equation}
	for the spatial profile, and 
\begin{equation}
\Sigma(R) = \frac{\rho_{0} b \pi}{ \Gamma(3\beta/2)} \left({1+ 
\frac{R^{2}}{b^{2}}}\right)^{(1-3\beta)/2} \Gamma( (3\beta- {1})/{2})
\end{equation}
for the projected profile.
	\item  A Sersic profile 
\begin{equation}
\Sigma(R)={\Sigma_{0}} \,\, {\exp\left({-{{({R\over a})}^{\nu 
}}}\right)} \,  ,
\end{equation}
for the projected profile, to which corresponds the spatial profile:~
\begin{equation}
\rho(r) =
{{{\Sigma_{0}}\, {\Gamma}({2\over {\nu }})}\over {2\,a\,{\Gamma}({{3 - p}\over {\nu }})}}
{\left(   {r\over a}\right)^{-p}}{\exp\left({-{{({r\over a})}^{\nu 
}}}\right)}
\end{equation}
where $p =1 -0.6097 \nu + 0.05463 \nu^2$ (Gerbal et al. 1997).
	\item A power-law for both the spatial and surface energy
density since the inverse Abel transform of a power law is a power law
with a power index decreased by one.
\end{enumerate} 

\subsection{ Non parametric analysis }
\label{s:np}
The non parametric inversion problem is concerned with finding the
best solution to \Eq{mass} for the cumulative mass profile when only
discretized and noisy measurements of $\Sigma$ and $\Sigma
\sigma_{p}^{2}$ are available (Gebhardt et al. 1996, Merritt 1996,
Pichon \& Thi\'ebaut 1998 and references therein).  In order to
achieve this goal these functions are written in some fairly general
form involving generically many more parameters than constraints and
such that each parameter controls only locally the shape of the
function.  The corresponding inversion problem is known to be
ill-conditioned: a small departure in the measured data (due to noise)
may produce drastically different solutions since these solutions are
dominated by artifacts due to the amplification of noise.  Some kind
of trade off must therefore be found between the level of smoothness
imposed on the solution in order to deal with these artefacts on the
one hand, and the level of fluctuations consistent with the amount of
information in the signal on the other hand.  Finding such a balance
is called the ``regularization'' of the inversion problem.

In practice the regularization can be imposed either directly upon the
projected model in data space or upon the unprojected model. The
latter (the non parametric inversion) is preferable for Abel
transforms such as \Eq{invAbel2} since the projection on the sky of a
given galaxy distribution is bound to be smoother than the galaxy
distribution itself.  Moreover, physical constraints such as
positivity of the galaxy distribution are also more stringent (and
better physically motivated) in model space.  Nevertheless it is
sometimes more straightforward to carry the regularization in model
space (a non parametric fit) and then carry the inversion numerically
when an explicit inversion formula such as \Eq{invAbel2} is
available.

Here, we apply both techniques to the recovery of the mass profile of 
ABCG~85.

\subsubsection{  Non parametric  fit }

We fit a regularized spline to $\log \Sigma$ and $\log \Sigma
\sigma_{p}^{2}$ as a function of $\log R$ with a linear penalty
function on the second derivative (i.e. which leaves invariant linear
functions of $\log_{10} R$ which are power laws of $R$).  We then make
explicit use of \Eq{invAbel2} to compute numerically $\rho$ and $\rho
\sigma_{r}^{2}$ together with their derivative.  Note that this
procedure is a non parametric fit rather than a non parametric
inversion, and the regularization parameter needs to be boosted to
account for the fact that the fit is then inverted to yield supposedly
smooth deprojected quantities.  In practice we use the regularization
parameter \(\mu_{0} = 5 \mu_{GCV} \) where $ \mu_{GCV} $ is given by
General Cross Validation as defined in the Appendix.

\subsubsection{ Non parametric inversion }

We fit the projection of a B-spline family which is sampled
logarithmically in radius with a linear penalty function on the second
derivative as discussed in the Appendix.  The coefficients of the fit
yield directly $\rho(r)$ and $\rho \sigma_r^2(r)$ which together with
\Eq{mass0} lead to the cumulative mass profile of ABCG~85.
As expected, the error bars on the corresponding mean profile are
larger for the non parametric inversion since this method imposes the
weakest prejudice on the expected mass profile.

\subsection{Results and discussion}

\begin{figure}
\centerline{\psfig{figure=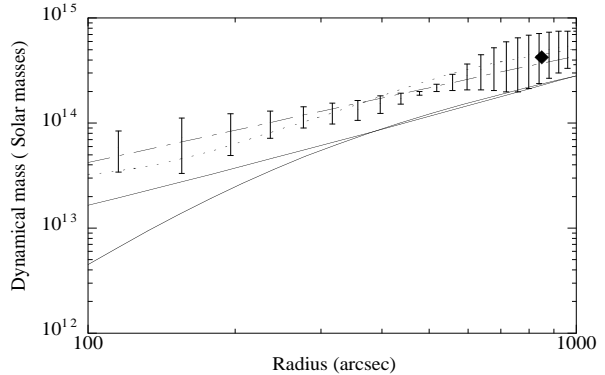,height=5cm,width=8cm}}
\caption[ ]{ Dynamical mass as  a  function of  radius derived with  various
methods.  Full lines: mass derived from X-ray data assuming a $\beta$--model
and a Sersic model for the gas distribution; small dotted line: complete non
parametric  inversion (i.e.  3D model   projected  in data space); the  mean
profile and the error  bars are estimated while  varying the binning from 11
to 18 neighbours; dot-dashed line: galaxy density and  pressure following a
power  law. The large square is  the dynamical  mass estimated with  the
Bahcall \& Tremaine method.}  
\protect\label{masses}
\end{figure}

We have obtained various dynamical integrated mass profiles, some of
which are displayed in Fig.~\ref{masses}, superimposed on those
obtained from X-ray data (Pislar et al., 1997). In order to avoid
having too many curves on this figure, we omitted the three curves
corresponding to the following cases: 2D non parametric fit, galaxy
density and pressure following $\beta$--models, and galaxy density
following a Sersic distribution and pressure following a
$\beta$--model; these three cases are almost indistinguishable from
the power law case. Notice that the limiting radius for the mass
obtained from optical data (\ODM) - 1000 arcsec - is smaller than the
limiting radius - 1300 arcsec - for the mass obtained from X-ray data
(\XDM); this is due to the lack of completeness of the galaxy velocity
catalogue in a larger region.

\begin{figure}
\centerline{\psfig{figure=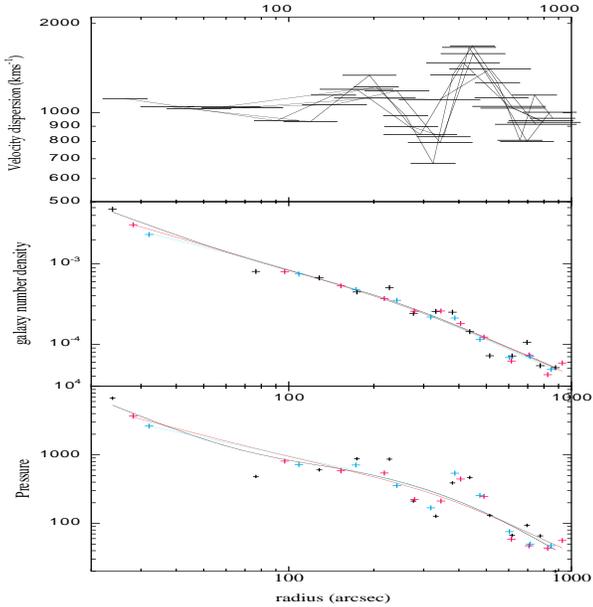,height=8cm,width=8cm}}
\caption[ ]{Top panel: velocity dispersion profile ; middle panel: density
profile; bottom panel: pressure profile. In the two bottom panels, the
observed points are indicated with crosses and typical fits with lines.
All these data were obtained with a floating binning mean method. The
horizontal lines show the corresponding bins.}
\protect\label{vedepr}
\end{figure}

\begin{figure}
\centerline{\psfig{figure=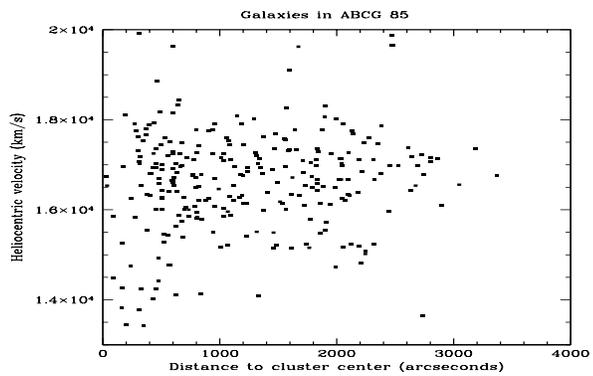,height=5cm,width=8cm}}
\caption[ ]{Velocity as a function of projected distance to the cluster 
center for galaxies in the cluster velocity range.}
\protect\label{vdall}
\end{figure}

We give in Fig.~\ref{vedepr} an example of a non-parametric fit of the
observed pressure (bottom panel) and observed profile (middle panel).
The observed points for the pressure are the result of the product of
the numerical profile with the velocity dispersion shown in the top
panel. This profile corresponds to the velocity distribution as a
function of projected distance to the cluster center displayed in
Fig.~\ref{vdall}.

One may notice that the various \ODM es are very close to one another,
the distances between the curves being smaller than the error bars
(see figure); the dynamical mass estimated with the Bahcall \&
Tremaine method is totally consistent with our estimate.

\subsubsection{`Optical' dynamical masses \textit{versus} \XDM es}

`Optical' dynamical masses are larger than \XDM es. At a distance of R=1000
arcsec, \ODM es  are $\simeq~4~10^{14}$ M$_\odot$, while \XDM es are
$\simeq~2~10^{14}$ M$_\odot$.

A simple explanation would be the following: from the spectral
capability of the ROSAT PSPC, Pislar et al. (1997) have derived an
\textsl{isothermal} plasma temperature of about 4 keV.  However
Markevitch et al. (1998) using ASCA have shown that in the centre of
ABCG~85 (where our analysis is performed) the temperature is about 8
keV. Such a discrepancy between ROSAT and ASCA determined temperatures
is not uncommon, since the energy range of ROSAT is lower than that of
ASCA, and therefore not well suited to measure cluster temperatures.
Since \XDM es are proportional to the temperature, the use of the
Markevitch et al. temperature would lead to a dynamical mass
$\simeq~4~10^{14}$ M$_\odot$ at R=1000 arcsec, equal to the \ODM es.
It is then tempting to conclude that the Markevitch et al.  high
plasma temperature is confirmed by the comparison of `optical' and
\XDM es. 

A careful analysis of the ABCG 85 temperature map provided by ASCA
(Markevitch et al. 1998) raises the question of the actual gas
temperature; the observed value of 8~keV is only valid for a region of
about 500 arcsec radius.

\begin{figure}
\centerline{\psfig{figure=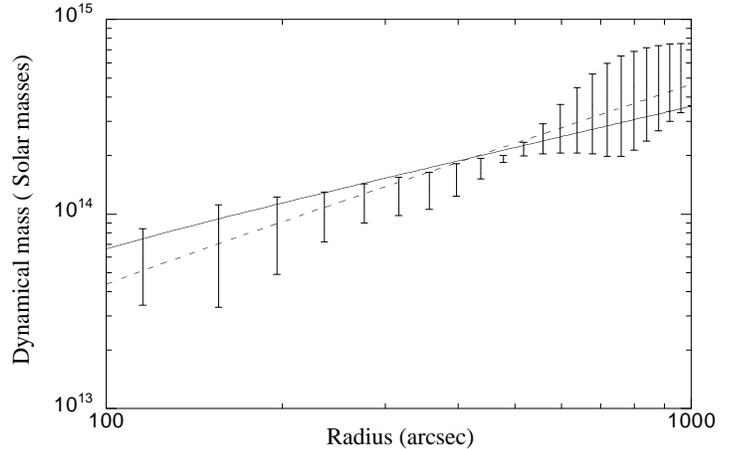,height=6cm}}
\caption[ ]{\XDM es estimated in two cases: dotted line: isothermal gas,
full line: non-isothermal gas; the error bars are the same as in 
Fig.~\ref{masses} for \ODM es.}
\protect\label{markevitch}
\end{figure}

A 3-D temperature profile must be defined by at least three
quantities, i.e. a slope (or equivalently an adiabatic index), a
length scale and a value of the temperature at a given radius. In the
case of ABCG 85, the data are too poor to recover the three above
values (Markevitch, private communication). However we have estimated
the dynamical mass assuming a ``mean'' value for the temperature slope
compatible with the whole sample of the Markevitch et al. data and
using the hydrostatic equation. Notice first that the asymptotic
behaviour for the integrated dynamical mass is (obviously) no longer
the isothermal one (i.e.  $M_{dyn}(r) \propto r^{-\gamma}$, with
$\gamma \sim 0.7$). As a consequence the amount of dark matter at
large scale would be largely reduced compared to the isothermal
behaviour.  However, we adress the question of the dynamical mass in a
region of only 1000 arcsec.  Using a value for the scale length of
order 500 kpc we are able to derive a M$_{dyn}$ profile; the obtained
mass compared to the isothermal one is displayed in
Fig.~\ref{markevitch}. As it is the case for some clusters analysed by
Markevitch et al. (private communication and poster at the Paris Texas
meeting) the non-isothermal mass is higher for small radii but
intersects the isothermal profile at a radius of $R_{cut} \sim 800$
arcsec. The resulting profile is clearly located in the region covered
with error bars as indicated in Fig.~\ref{markevitch}.

Our conclusion is that a non-isothermal analysis is not currently
possible due to the weakness of the temperature analysis at least for
ABCG 85, but is certainly a promising possibility in the future.

\begin{figure}
\centerline{\psfig{figure=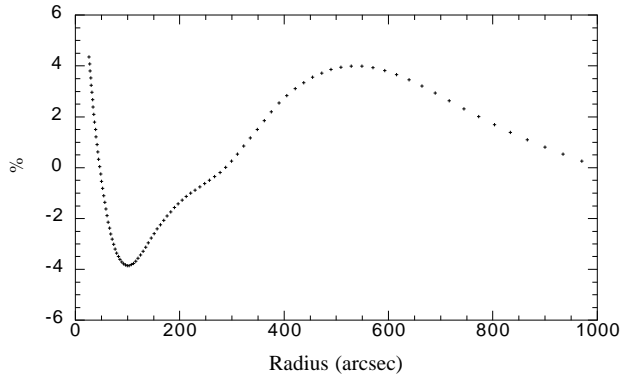,height=5cm}}
\caption[ ]{Residual of the deprojected velocity dispersion profile 
compared to a mean velocity dispersion of 1072 km$s^{-1}$.}
\protect\label{isoveloce}
\end{figure}

One may notice that \ODM es show the same rate of growth (i.e. $M(r)
\propto r$) as the \XDM es in the isothermal regime; the corresponding
densities vary as $r^{-2}$. We have calculated the 3-D galaxy velocity
dispersion profile (defined as the pressure to density ratio). The
mean velocity dispersion is 1072 km s$^{-1}$ and the residual of this
profile compared to this mean value is displayed in
Fig.~\ref{isoveloce}; the variation is $\leq 4 \%$ indicating that the
dispersion profile is constant with radius.

Therefore with similar hypotheses - isothermality for the X-ray
emitting plasma, isovelocity for the galaxies - \XDM es and \ODM es
are equal. This coherence validates the two independent techniques;
notice however that the static hypothesis is common for the two
methods. 

It is interesting to note that the dependence of the mass of the
different components with radius (between $\sim$100 and 1000 arcsec)
is: $M(r) \propto r^{0.2-0.3}$ for galaxies, $M(r) \propto r^{1.7}$
for the X-ray gas and $M(r) \propto r$ for dynamical matter, as noted
above in the isothermal case. 

\subsubsection{Additional comments}

The various modelling techniques implemented onto the masss profile of
ABCG 85 have led to very similar featureless powerlaw profiles for the
enclosed mass (Fig.~14). This property follows because even with a
sample of about 300 galaxies (all the galaxies in the cluster velocity
interval were included, i.e. 272 NoELGs and 33 ELGs) the inversion of
\Eq{Sigmau2} or \Ep{Sigma} is dominated by shot noise and requires
stringent regularisation.  As mentioned previously by Merritt \&
Tremblay (1994), we find here that a truly non parametric inversion
would require that nature provides many more galaxies per cluster.

We remind the reader that we have assumed an isotropic velocity
dispersion for the galaxies; the agreement found above shows that this
hypothesis is reasonable. It is easily accounted for by the fact that
the dominance of radial orbits is probably due to the continuous
infall of galaxies and small groups, as shown in particular for ABCG
85 (Durret et al. 1998b), which occurs mostly in the outskirts of the
cluster. It is only in the outer regions that the hydrostatic
hypothesis is also questionable.

A possible caveat is the following: only the velocity dispersion is
observed, while it is the pressure which is the important physical
quantity.  The estimate of the pressure as the numerical density
multiplied by the velocity dispersion is correct if there is no
equipartition between small and large galaxy masses, or expressed
differently if the velocity dispersion does not depend on the galaxy
mass.  We have checked in a central region of 750 arcsec radius that
the velocity dispersion does not depend on the magnitude; assuming a
constant mass to light ratio for all galaxies, this implies that the
velocity dispersion does not depend on galaxy mass.

\section{Conclusions}\label{discu}

The combined large field photographic plate and small field CCD
imaging catalogues, coupled with extensive spectroscopic data, have
led us to gather one of the largest amounts of data for a single
cluster.  These data have been used in the present paper to analyze
several properties of ABCG 85. Some of these properties have already
been discussed in the past by various authors (see Introduction), but
the large amount of data now available allows a more refined analysis,
leading either to derive new properties or to confirm previous results
with a high confidence level.

First, we have compared the distributions of emission line (ELGs) and
non-emission line galaxies (NoELGs), and shown that ELGs seem
intrinsically fainter than NoELGs, and do not appear as centrally
condensed as NoELGs, both spatially and in velocity space. ELGs show
an enhancement south of the nucleus, where groups are falling onto the
main cluster (as discussed in our previous paper by Durret et
al. 1998b).  This fits in well with the general view of this cluster:
the gas in the galaxies belonging to these groups is expected to be
shocked and consequently star formation should be more important in
the impact region at the epoch of actual galaxy infall, and less
important in the central regions of clusters where star formation
appears to be truncated. This has been shown to be the case for two
clusters at redshifts 0.2 and 0.4 by Abraham et al. (1996) and Morris
et al. (1998). Besides, the cluster analyzed by Abraham et al., ABCG
2390, shows evidence for a subcomponent infalling onto the main
cluster, as the south blob is falling onto ABCG 85.

Second, we have analyzed in detail the luminosity function of ABCG 85
in the R band, using a wavelet reconstruction technique. We have shown
with a high confidence level that a dip was present at an absolute
magnitude M$_{\rm R}\simeq -20.5$. This feature has also been detected
in several other clusters and can be accounted for by the
distributions of the various types of galaxies present in the cluster.
In this scenario, the dip would correspond to the separation between
elliptical and dwarf galaxies.

Third, parametric and non-parametric methods applied to our redshift
catalogue have allowed us to derive the dynamical properties of the
cluster.  We find that the dynamical mass profiles derived from the
X-ray gas and galaxy distributions agree if the temperature of the
X-ray emitting plasma is about 8 keV. Between 250 and 1000 arcsec,
whatever technique we apply (parametric or not), and whatever data we
use (X-ray or optical), the slopes of the dynamical mass profiles are
the same ($M(r) \propto r$). In this region, both the X-ray plasma and
the ``gas'' of galaxies are isothermal, and the galaxy velocity
dispersion is isotropic. If we take into account the temperature
gradient of the X-ray gas, the dynamical mass is reduced. If we take
into account a possible temperature gradient of the X-ray gas, the
\XDM\ is reduced at very large scale but is still comparable to the
\ODM\ in the X-ray emitting region.

Although this paper is the last one of the series on ABCG 85, the
analysis of a much larger area in that region of the sky is planned in
a near future: we have recently obtained about 300 new redshifts in
the direction of ABCG 87 (in collaboration with M.~Colless), and have
the project of obtaining redshifts for the various clusters and groups
aligned along PA$\sim 160^\circ$ and in which ABCG 85 seems
embedded. We also intend to discuss the large scale structure
properties of the universe in the direction of ABCG 85, based on the
large scale velocity features of our velocity catalogue.

\begin{acknowledgements}
We are very grateful to Val\'erie de Lapparent and H. Lin for making
their counts in the R band available to us, and to A. Biviano for his
adaptive kernel program.  We acknowledge interesting discussions with
G.B.  Lima-Neto.  CP would like to thank E.~Thi\'ebaut for many
valuable discussions.  Last but not least, we gratefully thank Dario
Fadda for his help in tackling the problem of estimating error bars on
the luminosity function and Fr\'ed\'eric Magnard for his help in these
computations. We appreciated several interesting suggestions by the
referee, A. Biviano. C.L. acknowledges financial support by the CNAA
(Italia) fellowship reference D.D. n.37 08/10/1997, and
C.P. acknowledges funding from the Swiss NF.
\end{acknowledgements}

\appendix

\section{ Non parametric analysis}

\begin{figure}
\centerline{\psfig{figure=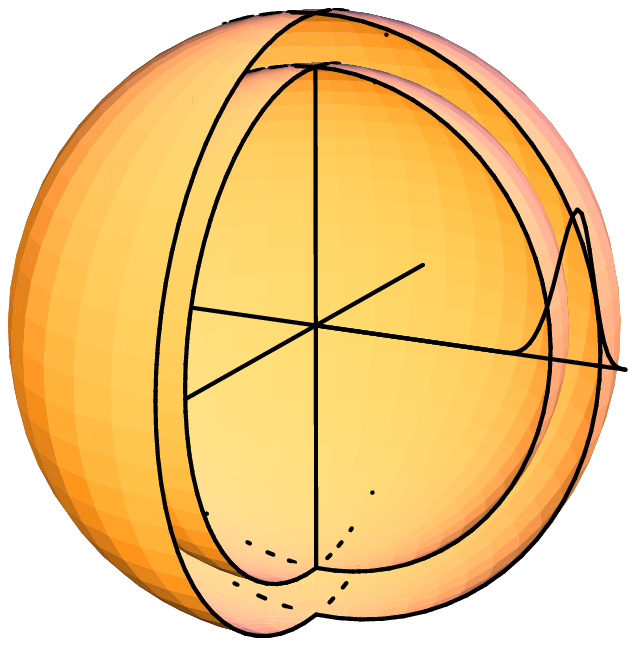,height=6cm}}
\vskip -1.0truecm
\centerline{\psfig{figure=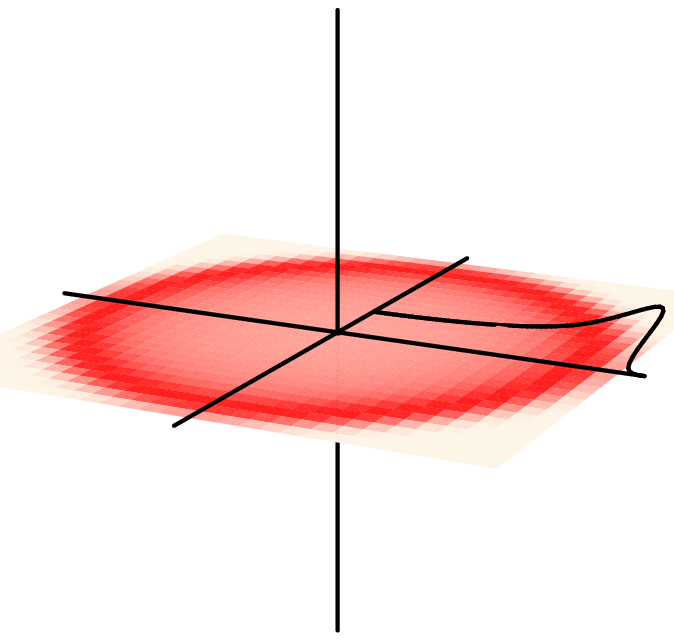,height=4cm}}
\vskip -0.8truecm
\caption{Sketch of  deprojection: concentric  B-spline shells of  increasing
logarithmic radii   are projected onto  the  sky; their  corresponding light
distribution is calculated analytically and  represents the basis over which
the data is expanded.}
\end{figure}

\begin{figure}
\centerline{\psfig{figure=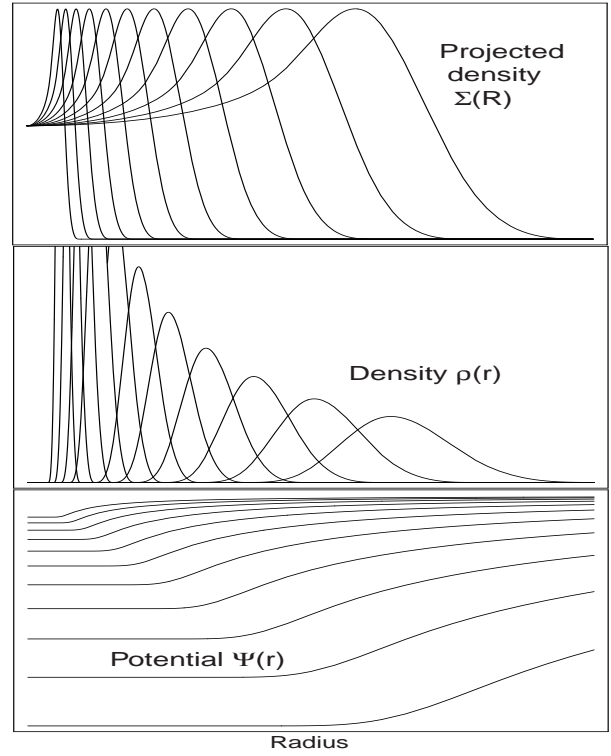,height=10cm,width=8cm}}
\caption[ ]{The B-spline  basis  functions and their  transform as a
function of radius.   Top panel: the B-spline projection; middle panel: 
corresponding density distribution; bottom panel: corresponding  self 
consistent potential.}
\end{figure}

The  non parametric   solutions  of \Eq{Sigma}   and  \Ep{Sigmau2} are
described  by their  projection onto   a complete  basis of  functions
$$\{e_k(r)\}_{ k=1,\ldots,n}$$ of finite support,  which are chosen here
to be cubic B-splines (i.e. the unique $C^2$ function
 which is defined to be a cubic over 4 adjacent intervals and zero
outside, with the extra property that it integrates to unity  over
that interval):
\begin{equation}
\rho(r) = \sum_{k=1}^{n}  \rho_{k} e_{k}(r) 
  , 
\end{equation}
\begin{equation}
  \rho \sigma_{r}^{2}(r)  =      \sum_{k=1}^{n} (\rho
\sigma_{r}^{2})_{k} e_{k}(r)
\, .
\end{equation}
The parameters  to  fit  are  the  weights $\rho_k$ and
$(\rho \sigma_{r}^{2})_{k}$.
Calling  $\M{x}=\{ \rho_{k}\}_{k=1,..n}$ or $\{(\rho \sigma_{r}^{2})_{k}\}_{k=1,..n}$
(the   parameters)  and $\tilde{\M{y}}=\{  \Sigma_{k}\}_{k=1,..K}$  or   $\{
(\Sigma \sigma_{p}^{2})_{k}\}_{k=1,..K}$ (the  measurements)   \Eq{Sigma} and  \Ep{epsilon}
then become formally
\begin{equation}
\tilde{\M{y}}= \M{a} \cdot \M{x} \, ,
\end{equation}
where $\M{a}$ is a $K\times n$ matrix with entries given by
\begin{equation}
 a_{i,k}= 2 \, 
\int_{R_{i}}^\infty e_{k}(r) \frac{r  \d  r  }{\sqrt{r^2-R_{i}^{2}}} =
{\cal A}_{R_{i}}(e_{k}) \, .  \EQN{eqn}
\end{equation}

The projection and the self consistent potential of B-splines can be
computed analytically.  Their knots can be placed arbitrarily in order
to resolve high frequencies in the profiles which are believed to be
signal rather than noise (this is a requirement when using a penalty
function which operates on the spline coefficient since imposing a
correlation between these coefficients would truncate the high
frequency).  The analytic properties of B-splines and their transform
turns out handy in particular since Taylor expansions are available
when dealing with exponential profile where the dynamical range is
large.  Another useful property of B-spline is extrapolation: the
correlation of the spline coefficient induced by the penalty function
yields an estimate for the behaviour of the profile beyond the last
measured point; since the Abel transform requires integration to
infinity, this estimate corrects in part for the truncation. Note
that an explicit analytic continuation of the model can be added to
the spline basis if required.  Finally here the requirement is that
$\M{x}$ is smooth, which is more strigent than requiring that $\Sigma$
(or $\Sigma \sigma_{p}^{2}$) are smooth.

Assuming that we have access to  discrete measurements of $\Sigma$ and
$\Sigma \sigma_{p}^{2}$ (via binning  as discussed above), and that the
noise in $\Sigma$ and $\Sigma \sigma_{p}^{2}$  can be considered to be
Normal,  we can estimate the  error  between the measured profiles and
the non parametric B-spline model as
\begin{equation}
\R{L}_\R{}(\M{x}) =\chi^2(\M{x}) = 
		\T{({\tilde{\M{y}}} - \M{a}\mdot \M{x})} \mdot \M{W}
		\mdot ({\tilde{\M{y}}} - \M{a} \mdot \M{x}) \,, \EQN{Lquad}
\end{equation}  where 
 the weight matrix  $\M{W}$ is the inverse of  the covariance matrix  of the
data (which is diagonal for uncorrelated noise with diagonal elements equal
to one over the data variance).
Linear penalty functions obey
 \begin{equation}
 	\R{R}_\R{}(\M{x}) = \T{\M{x}} \mdot \M{K} \mdot \M{x}\,, \EQN{Pquad}
 \end{equation} where $\M{K}$ is a positive definite matrix.
In practice, we use $\M{K} = \T{\M{D}}\cdot \M{D}$ where $\M{D}$
 is a finite difference second order operator
\begin{equation}
	\M{D}= {\rm Diag}_{3}[-1,2,-1] \equiv  \left[\begin{array}{cccccc}
-1 & 2 & -1 & 0 &\ldots & 0\\
0 & -1 &  2 & -1 & \ldots &  \\
& & \ddots & \ddots &  \ddots &\\
& &  & & &\\
\end{array}\right] \, .
\end{equation}

In   short, the  solution   of \Eq{Sigma} (or \Eq{Sigmau2})  is found by
minimizing the  quantity  $Q(\M{x})=L(\M{x})+\mu\,R(\M{x})$ where $L(\M{x})$
and $R(\M{x})$ are  respectively  the likelihood and   regularization terms
given by \Eq{Lquad} and \Ep{Pquad},
$\M{x}$  are  the  (large  number) of  parameters,  and  where the  Lagrange
multiplier  $\mu>0$  allows us  to tune   the level of  regularization. The
introduction of the  Lagrange multiplier $\mu$ is  formally justified by the
fact that  we want to minimize  $Q(\M{x})$,  subject to the  constraint that
$L(\M{x})$   should  be   equal   to  some   value.    For   instance,  with
$L(\M{x})=\chi^2(\M{x})$ the problem is   to minimize $Q(\M{x})$ subject  to
the  constraints      that      $L(\M{x})$      is     in     the      range
$N_\R{data}\pm\sqrt{2\,N_\R{data}}$). In practice, the minimum of
\begin{equation}
	Q_\R{}(\M{x}) =
		\T{({\tilde{\M{y}}} - \M{a}\mdot \M{x})} \mdot \M{W}
		\mdot ({\tilde{\M{y}}} - \M{a} \mdot \M{x})
		+ \mu \, \T{\M{x}} \mdot \M{K} \mdot \M{x} \, 
	\label{e:Q-quad}
\end{equation}
\noindent   is:
\begin{equation}
	\M{x}_\R{} = (\T{\M{a}} \mdot \M{W} \mdot \M{a} + \mu\,\M{K})^{-1}
		\mdot \T{\M{a}} \mdot \M{W} \mdot \tilde{\M{y}}\,.
		\label{e:Q-quad-solution} 
\end{equation}

The last remaining issue involves setting the level of regularization.
The  so-called cross-validation method  (Wahba 1990)  adjusts the value  of
$\mu$ so as to minimise residuals between the  data and the prediction
derived from the data.  Let us define
\begin{equation}
	\tilde{\M{a}}(\mu) =\M{a} \mdot   (\T{\M{a}} \mdot  \M{W}   \mdot \M{a}  +
		\mu\,\M{K})^{-1}   \mdot  \T{\M{a}}     \mdot  \M{W}
\,.  \label{e:Q-quad-solution}
\end{equation}
We make use of the value for $\mu$  given  by Generalised Cross Validation 
(GCV) (Wahba \& Wendelberger 1979) estimator  corresponding to the mimimum of
\begin{equation}
\mu_0 \equiv GCV(\mu) = {\rm min}_\mu\left\{
\frac{||( \M{1}-	\tilde{\M{a}}) \mdot \tilde{\M{y}} ||^2}{
\left[{\rm trace}( \M{1}-	\tilde{\M{a}}) \right]^2} \right\} \,.
  \EQN{GCV}
\end{equation}

\end{document}